\documentclass{llncs}
\usepackage{graphicx}

\usepackage{booktabs} 
\usepackage{algpseudocode,algorithm,algorithmicx}
\usepackage{enumerate}
\usepackage{skmath}
\usepackage{mathtools}
\usepackage{multirow}
\usepackage{xspace}
\usepackage{tikz}
\usetikzlibrary{arrows,calc,positioning,fit,shapes.multipart,automata}
\usepackage{subcaption}
\usepackage{pgfplots,pgfplotstable}
\usepackage{url}
\usepackage{listings}
\usepackage{xcolor}
\usepackage{array}

\definecolor{dkgreen}{rgb}{0,0.6,0}
\definecolor{gray}{rgb}{0.5,0.5,0.5}
\definecolor{mauve}{rgb}{0.58,0,0.82}
\definecolor{verylghbl}{RGB}{237,240,244}

\def\ie{\emph{i.e.}\xspace}	
\def\eg{\emph{e.g.}\xspace} 
\newcolumntype{H}{>{\setbox0=\hbox\bgroup}c<{\egroup}@{}}
\newcommand{\unaryminus}{\scalebox{0.5}[1.0]{\( - \)}}
\newcommand{\expnum}[2]{{#1}\mathrm{e}{#2}}
\newcommand{\nexpnum}[2]{{#1}\mathrm{e}{\unaryminus #2}}

\algrenewcommand\algorithmicrequire{\textbf{Input:}}
\algrenewcommand\algorithmicensure{\textbf{Output:}}
\algnewcommand{\LineComment}[1]{\Statex \(\triangleright\) #1}
\newcommand{\pluseq}{\mathbin{{+}{=}}}

\algblock{ParFor}{EndParFor}
\algnewcommand\algorithmicparfor{\textbf{parallel for}}
\algnewcommand\algorithmicpardo{\textbf{do}}
\algnewcommand\algorithmicendparfor{\textbf{end\ for}}
\algrenewtext{ParFor}[1]{\algorithmicparfor\ #1\ \algorithmicpardo}
\algrenewtext{EndParFor}{\algorithmicendparfor}

\pgfplotsset{
	discard if not/.style 2 args={
		x filter/.code={
			\edef\tempa{\thisrow{#1}}
			\edef\tempb{#2}
			\ifx\tempa\tempb
			\else
			\def\pgfmathresult{inf}
			\fi
		}
	}
}

\pgfplotsset{compat=1.12}
\tikzset{adjust near node color/.code={%
		\pgfplotscolormapdefinemappedcolor\pgfplotspointmetatransformed%
		\definecolor{mapped node color}{rgb}{\pgfmathresult}%
		\pgfkeys{/tikz/text=mapped node color!80!black}%
	}
}

\def\myplotmark#1{%
	\begin{pgfpicture}\pgfuseplotmark{#1}\end{pgfpicture}%
}

\usepgfplotslibrary{colorbrewer}
\pgfplotsset{cycle list/Dark2}

\newcommand{\method}{\textproc }
\newcommand{\variable}{\textit }
\def\maxBlockSize{\texttt{MAX\_BLOCK\_SIZE}\xspace}
\def\maxSimilarity{\texttt{MAX\_SIMILARITY}\xspace}
\def\maxKeysPerRecord{\texttt{MAX\_KEYS}\xspace}

\newcommand{\repeatthanks}{\textsuperscript{\thefootnote}}

\begin{document}
\title{Scalable Blocking for Very Large Databases}

\author{Andrew Borthwick\thanks{Equal Contribution} \and 
	Stephen Ash\repeatthanks \and 
	Bin Pang \and 
	Shehzad Qureshi \and
	Timothy Jones
}
\institute{AWS AI Labs \email{\{andborth,ashstep,pangbin,shezq,timjons\}@amazon.com}}

\maketitle

\begin{abstract}
In the field of database deduplication, the goal is to find approximately matching records within a database. \textit{Blocking} is a typical stage in this process that involves cheaply finding candidate pairs of records that are potential matches for further processing.
We present here \textit{Hashed Dynamic Blocking}, a new approach to blocking designed to address datasets larger than those studied in most prior work. Hashed Dynamic Blocking (HDB) extends Dynamic Blocking, which leverages the insight that rare matching values and rare intersections of values are predictive of a matching relationship. We also present a novel use of Locality Sensitive Hashing (LSH) to build blocking key values for huge databases with a convenient configuration to control the trade-off between precision and recall. HDB achieves massive scale by minimizing data movement, using compact block representation, and greedily pruning ineffective candidate blocks using a Count-min Sketch approximate counting data structure.  
We benchmark the algorithm by focusing on real-world datasets in excess of one million rows, demonstrating that the algorithm displays linear time complexity scaling in this range. Furthermore, we execute HDB on a 530 million row industrial dataset, detecting 68 billion candidate pairs in less than three hours at a cost of \$307 on a major cloud service. 

\end{abstract}


\section{Introduction}
\label{introduction}
Finding approximately matching records is an important and well-studied problem \cite{Elmagarmid2007}. The challenge is to identify records which represent the same real-world entity (e.g. the same person, product, business, movie, etc.) despite the fact that the corresponding data records may differ due to various errors, omissions, or different ways of representing the same information.%
%
For many database deduplication/record linkage applications a common approach is to divide the problem into four stages \cite{koudas2006record}: Normalization \cite{ash2015embracing}, Blocking \cite{Christen2011,McNeill2012,Papadakis2016}, Pairwise Matching \cite{Mudgal2018,Chen2011}, and Graph Partitioning \cite{Hassanzadeh2009,Reas2018}.

This work focuses on the problem of blocking very large databases with record counts between 1M and 530M records.  We focus on databases in this range due to their importance in industrial settings. We make a case that different blocking algorithms will be successful in this range than are effective on databases with fewer than 1 million records. 
The experimental results that we show on real-world datasets with 50M or more records, in particular, are unusual in the literature.

In contrast to prior work such as \cite{Bilenko2006} which seeks to build an optimal set of fields on which to block records, the philosophy of the dynamic blocking family of algorithms \cite{McNeill2012} is to avoid selecting a rigid set of fields and instead dynamically pick particular \textit{values} or combinations of values on which to block. As an example of blocking on a fixed, static set of blocking key fields, consider a system to deduplicate U.S. person records by simply proposing all pairs of persons who match on the field \variable{last\_name}. This would be prohibitively expensive due to the necessity of executing a pairwise matching algorithm on each of the $\binom{1,400,000}{2}$ pairs of people who share the surname ``Jones''. A pairwise scoring model averaging 50~$\mu$secs would take $\approx567$ days to compare all ``Jones'' pairs.  

On the other hand, suppose that we statically select the pair of fields \variable{(first\_name, last\_name)} as a single blocking key.
This solves the problem of too many ``Jones'' records to compare, but is an unfortunate choice for someone with the name ``Laurence Fishburne'' or ``Shehzad Qureshi''. Both of these surnames are rare in the U.S. A static blocking strategy which required both given name and surname to match would risk missing the pair (``laurence fishburne'',``larry fishburne'') or (``shehzad  qureshi'',``shezad qureshi'').  
Differentiating between common and less common field/value pairs in the blocking stage fits with the intuition that it is more likely that two records with the surname ``Fishburne'' or ``Qureshi'' represent the same real-world individual than is the case for two records with the surname ``Jones'', which is an intuition backed up by algorithms that weight matches on rare values more strongly than matches on common values \cite{Chen2011,koudas2006record,wang2014probabilistic}.  



This work makes the following contributions: (1) We describe a new algorithm called Hashed Dynamic Blocking (HDB) based on the same underlying principle as dynamic blocking \cite{McNeill2012}, but achieves massive scale by minimizing data movement, using compact block representation, and greedily pruning ineffective candidate blocks. We provide benchmarks that show the advantages of this approach to blocking over competing approaches on huge real-world databases. (2) Our experimental evidence emphasizes very large real-world datasets in the range of 1M to 530M records. We highlight the computational complexity challenges that come with working at this scale and we demonstrate that some widely cited algorithms break down completely at the high end. (3) We describe a version of Locality Sensitive Hashing applied to \textit{blocking} that is easily tunable for increased precision or increased recall. Our application of LSH can generate (possibly overlapping) blocking keys for multiple columns simultaneously, and we provide empirical evaluation of LSH versus Token Blocking to highlight the trade-offs and scaling properties of both approaches.

\section{LSH and block building}
\label{topLevelBlocking}
Like most other blocking approaches such as Meta-blocking \cite{Papadakis2014}, dynamic blocking begins with a set of records and a \textit{block building} step that computes a set of \textit{top-level} blocks, which is a set of records that share a value computed by a block blocking process, $t$, where $t$ is a function that returns a set of one or more \textit{blocking keys} when applied to an attribute $a_k$ of a single record, $r$. The core HDB algorithm described in Section~\ref{fdb_alogorithm_section} is agnostic to the approach to block building.  

With structured records, one can use domain knowledge or algorithms to pick which block building process to apply to each attribute. We term \textit{Identity Block Building} as the process of simply hashing the normalized (\eg lower-casing, etc.) attribute value concatenated to the attribute id to produce a blocking key. Thus the string ``foo'' in two different attributes returns two different top-level blocking keys (\ie hash values). For attributes where we wish to allow fuzzier matches to still block together, we propose \textit{LSH Block Building} as described in the next section. Alternatively, \textit{Token Blocking} \cite{Papadakis2014} is a schema-agnostic block building process where every token for every attribute becomes a top-level blocking key. Note that, unlike Identity Blocking Building, the token ``foo'' in two different attributes will return just a single blocking key.

\subsection{LSH block building}
\label{lsh_support}

\begin{figure*}[t]
    \centering
    \begin{subfigure}[t]{0.45\textwidth}

		\resizebox{\linewidth}{!}{
		\begin{tikzpicture}
		  \begin{axis}[ymin=0.0
		  	,ymax=1.0
		  	,xmin=0.0
		  	,xmax=1.0
		  	,ylabel={Probability of Blocking}
		  	,xlabel={Jaccard Similarity}
		  	,no marks
		  	,grid=both
		  	,legend style={font=\scriptsize,at={(0.02,0.98)},anchor=north west}
		  	]
		  	\addplot+[discard if not={lshaxis}{1t1}] table[x=jac, y=prob, col sep=comma] {data/lsh-jaccard.csv};
		  	\addlegendentry{$LSH(1,1)$}
		  	\addplot+[discard if not={lshaxis}{3t8}] table[x=jac, y=prob, col sep=comma] {data/lsh-jaccard.csv};
		  	\addlegendentry{$LSH(3,8)$}
		  	\addplot+[discard if not={lshaxis}{6t7}] table[x=jac, y=prob, col sep=comma] {data/lsh-jaccard.csv};
		  	\addlegendentry{$LSH(6,7)$}
		    \addplot+[discard if not={lshaxis}{10t6}] table[x=jac, y=prob, col sep=comma] {data/lsh-jaccard.csv};
		    \addlegendentry{$LSH(10,6)$}
		    \addplot+[discard if not={lshaxis}{12t5}] table[x=jac, y=prob, col sep=comma] {data/lsh-jaccard.csv};
		    \addlegendentry{$LSH(12,5)$}
		    \addplot+[discard if not={lshaxis}{14t4}] table[x=jac, y=prob, col sep=comma] {data/lsh-jaccard.csv};
		    \addlegendentry{$LSH(14,4)$}
			\addplot+[discard if not={lshaxis}{16t3}] table[x=jac, y=prob, col sep=comma] {data/lsh-jaccard.csv};
			\addlegendentry{$LSH(16,3)$}
		  \end{axis}
		\end{tikzpicture}
		}
        \caption{Probability of blocking vs Jaccard of the record pair's attribute under various $LSH(b,w)$ settings}
        \label{lshJaccardGraph}
    \end{subfigure}
    \quad
    \begin{subfigure}[t]{0.45\textwidth}
		
		\resizebox{\linewidth}{!}{
		\begin{tikzpicture}
		  \begin{axis}[ymin=0.00025
		  	,ymax=0.0075
		  	,xmin=0.3
		  	,xmax=1.0
		  	,enlargelimits=false
		  	,ylabel={$PQ$ (log-scale)}
		  	,xlabel={$PC$}
		  	,axis y line*=left
		  	,ylabel near ticks
		  	,yminorticks=true
		  	,ymode=log
       		,log basis y={10}
       		,grid=both
       		,visualization depends on={\thisrow{baspt} \as \perpointmarksize}
            ,scatter/@pre marker code/.append style={
                /tikz/mark size=\perpointmarksize,
            }
       		,visualization depends on={value \thisrow{lsh} \as \labellsh}
       		,visualization depends on={value \thisrow{pos} \as \myPos}
    		,nodes near coords align={horizontal}
		  	]
		    \addplot+[only marks
		    	,scatter
		    	,mark=* 
		    	,opacity=0.7
		    	,point meta=explicit
		    	,discard if not={include}{1}
		    	,nodes near coords*={\labellsh}
		    	,colormap={mycolormap}{rgb255=(228,26,28); rgb255=(55,126,184); rgb255=(77,175,74);
		    	rgb255=(152,78,163);rgb255=(255,127,0);rgb255=(55,126,184);rgb255=(153,153,153)}
		    	,every node near coord/.append style={
                	font=\scriptsize,
                	adjust near node color,
                	opacity=1.0,
                	\myPos=\perpointmarksize pt - 1.75 pt
            	}
		    	] table[x=PC, y=PQ, col sep=ampersand, meta=wid] {data/lsh-scholar-scatter3.csv};
		   \end{axis}
		\end{tikzpicture}
		}
        \caption{PQ vs PC on SCHOLAR under various $\text{LSH}(b,w)$ settings. The diameter of the point relates to the number of pairs produced.}
        \label{lshResultsScholar}
    \end{subfigure}
    
    \caption{$PQ$ and $PC$ of various LSH settings on the SCHOLAR dataset}
    \label{lshScholarScatter}
\end{figure*}
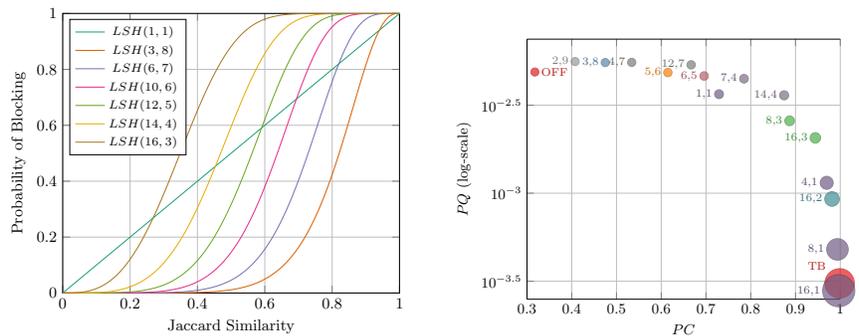

In this work we propose a new block building approach which incorporates Locality Sensitive Hashing (LSH) \cite{indyk1998approximate,Gionis1999} with configurable parameters to control the precision, recall trade-off per column. LSH block building creates multiple sets of keys that are designed to group similar attribute values into the same block. We leverage a version of the algorithm which looks for documents with high degrees of Jaccard similarity \cite{Leskovec2014,VanDam2016} of tokens. Here a \textit{token} could be defined as a \textit{word}, a word $n$-gram, or as a character $q$-gram. 

\cite{Leskovec2014} describes a method in which, for each document $d$, we first apply a minhash algorithm \cite{broder2000min} to yield $m$ minhashes and we then group these minhashes into $b$ \textit{bands} where each band consists of $w = m/b$ minhashes. In our approach each of these bands constitutes a blocking key.  Now consider a function $LSH(b,w,j)$, in which $b$ and $w$ are the LSH parameters mentioned above and $j$ is the Jaccard similarity of a pair of records, then $LSH(b,w,j)$ is the probability that the attributes of two records with Jaccard similarity of $j$ will share at least one key and can be computed as:
%
$LSH(b,w,j) = 1 - {(1 - {j^w}   )     }^b$
%
$LSH(b,w,\cdot)$ has an attractive property in that the probability of sharing a key is very low for low Jaccard similarity and very high for high Jaccard similarity.  Figure~\ref{lshScholarScatter} graphs $LSH(b,w,\cdot)$ for various values of $(b,w)$, which gives us a range of attractive trade-offs on the Pair-Quality (\ie precision) versus Pair-Completeness (\ie recall) curve by varying the two parameters for LSH, $b$ and $w$.

\subsection{Prior Work on Block Building}
Our block building techniques are strongly distinguished from prior work in the field.  
For example, \cite{Chu2016} makes an assumption that the block-building phase yields a strict partitioning of the database, although multiple distinct passes can be used to reduce false negatives \cite{Chu2016,Elmagarmid2007}.  Our approach, by contrast leverages the fact that the block building strategies discussed in Section~\ref{topLevelBlocking} yield blocks that are, by design, highly overlapping and all the blocks are processed in a single pass.  Another distinction is that \cite{Chu2016} generates a single minHash on a single attribute from which it builds blocks, which in terms of our approach would correspond to using a degenerate LSH with the parameters of $b=1, w=1$ on only a single column.  
Figure~\ref{lshScholarScatter} includes this $LSH(1,1)$ configuration, which highlights its particular point in the precision, recall curve for the SCHOLAR dataset. In this case, our $LSH(14,4)$ improves recall by $\approx20\%$ with only $\approx1\%$ change in precision.

Use of LSH is fairly common in the literature. \cite{papadakes2019} has an extensive recent survey.  \cite{Leskovec2014} has a useful tutorial on LSH that uses deduplication as an example.  However, our approach is new in that we do not apply LSH to the record as a whole, but rather LSH is applied selectively to columns where it makes sense (\eg columns consisting of multi-token text), while columns consisting of scalar values generate top-level blocks through trivial identity block building.
Using the nomenclature of a recent survey on blocking \cite{papadakes2019}, our block building strategy yields top-level blocks which are neither \textit{redundancy-free}, where every entity is assigned to exactly one block, nor is it \textit{redundancy-positive}, in which every entity is assigned to multiple blocks, such that the ``more blocks two entities share, the more similar their profiles are likely to be''. The latter is due to the fact that by construction LSH bands act as redundant blocking keys that do connote some similarity (\eg higher jaccard) but less similarity than co-occurring non-LSH keys.

\section{Hashed Dynamic Blocking}
\label{fdb_alogorithm_section}
\label{algorithmOverview}

\begin{figure*}[t]
	\centering
	\fbox{\includegraphics[clip, trim=0.4cm 8.5in 1.7in 0in, width=0.9\textwidth]{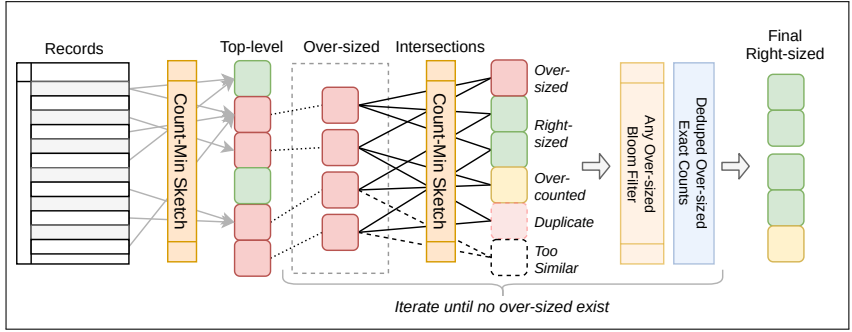}}
	\caption{Diagram illustrating how candidate blocks are processed in Hashed Dynamic Blocking}
	\label{fig:hdbDiagram}
\end{figure*}
%
%
%
Dynamic blocking \cite{McNeill2012} takes the approach of \textit{finding} overlapping subsets of records that share enough blocking key values in common to make the size of each subset small. We impose this subset size threshold as a way to balance the precision and recall of each emitted subset of records.   
%
%
%
At each iteration of the algorithm, we partition the blocking keys between those which are above and below this threshold: \maxBlockSize. Those below the max block size are deemed to be \textit{accepted} or right-sized.  For all accepted blocks $b_i$, the blocking phase is over. That is, later in the pairwise matching phase, we will compare all $\binom{|b_i|}{2}$ pairs of records from all right-sized blocks.  However, for over-size blocks, we need to find additional co-occurring blocking key values. 
To do this we compute the logical intersections of all pairs of over-size blocks. For example, if the ``Jones'' block described in Section \ref{introduction} had more than \maxBlockSize records, it could be intersected with other blocks generated by values that also had more than \maxBlockSize records such as the block for the first\_name ``Tim".  The logical intersection of these two blocks may be under the threshold. We progressively intersect blocking key \textit{values} in this way until we have no over-size blocks left. As described below, we use a few heuristics to guide this search for right-sized intersections to make this converge efficiently.

\subsection{Algorithm detailed description}

\begin{algorithm}[t]
	\caption{Hashed Dynamic Blocking}
	\label{alg:hashedDynBlockingTopLevel}
	\begin{algorithmic}[1]
		\Require{$R$, a dataset of records, $r$, to be blocked, with each record having a long identifier, $rid$}
		\Ensure{a \textit{blocked} dataset of deduplicated record pairs}
		\Function{HashedDynamicBlocking}{$R$}

		\State $K \gets \Call{BlockOnKeys}{R}$ 
		
		\State $(K_R, \tilde{K}_O) \gets \Call{RoughOversizeDetection}{K}$ \label{alg:hdbTopLoopStart}
		\State $(\hat{K}_R, K_O) \gets \Call{ExactlyCountAndDedupe}{\tilde{K}_O}$
		\State $K_R \gets K_R \cup \hat{K}_R$
		
		\While{$K_O \ne \emptyset$}
			\State $K' \gets \Call{IntersectKeys}{K_O}$
			\State $(K'_R, \tilde{K}'_O) \gets \Call{RoughOversizeDetection}{K'}$
			\State $(\hat{K}'_R, K'_O) \gets \Call{ExactlyCountAndDedupe}{\tilde{K}'_O}$
			\State $K_R \gets K_R \cup \hat{K}'_R$ \label{alg:unionCorrectedBlocks}
			\State $K_O \gets K'_O$
		\EndWhile \label{alg:hdbTopLoopEnd}
		
		\State \Return{$\Call{RemoveDupePairs}{K_R}$}
		\EndFunction
	\end{algorithmic}
\end{algorithm}

Algorithm~\ref{alg:hashedDynBlockingTopLevel} describes the high level algorithm of Hashed Dynamic Blocking (HDB). HDB is implemented in Apache Spark\footnote{https://spark.apache.org/}. For clarity we show the pseudocode in an imperative style, but in the implementation everything is implemented as a sequence of lazy map and reduce operations. In particular, we use the keyword \textbf{parallel for} to indicate which loops are actually \texttt{map} operations (as in the MapReduce paradigm).

In \method{HashedDynamicBlocking}, first, the function \method{BlockOnKeys} applies the configured block building functions to the input dataset, $R$, as described in Section~\ref{topLevelBlocking}. The resulting dataset, $K$, is an inverted index of blocking keys for each record ID. It is important to note that we do \textit{not} maintain or materialize the \textit{flipped} view of block key to list of record IDs in that block during these iterations. Materializing this view is relatively expensive as it requires shuffling the data across the cluster, and thus we only do it once at the end after all of the right-sized blocks have been determined. HDB operates exclusively on 64-bit record IDs and a sequence of hashed blocking keys represented by 128-bit hash values derived from the top-level blocking attribute values. After top-level block building, the actual attributes of the records are no longer needed and do not flow through the algorithm.

Figure~\ref{fig:hdbDiagram} visualizes the logical steps of how we identify right-sized and over-sized blocks. In each iteration, we lazily intersect the previously identified over-size blocking keys. We use fast approximate counting in \method{RoughOversizeDetection} (Algorithm \ref{alg:roughOversizeDetection}) to quickly identify which of these are right-sized blocks, $K_R$, and possibly over-sized blocks, $\tilde{K}_O$. 

Our probabilistic counting data structure \textit{might} over-count, and therefore some of the identified blocks in $\tilde{K}_O$ may in fact be right-sized. The function \method{ExactlyCountAndDedupe} (Algorithm \ref{alg:exactCountAndDedupe}) post-processes $\tilde{K}_O$ to accurately identify any over-counted right-sized blocks, $\hat{K}_R$. This function also de-duplicates the over-sized blocks efficiently, resulting in the final true, unique set of over-sized blocks left over, $K_O$, for the next iteration. After all iterations, the set of resultant right-sized blocks may have some duplicate pairs. Thus in function \method{RemoveDupePairs}, we remove duplicate pairs as described in Section~\ref{pairDedupe}.



\subsubsection{Intersecting Keys}

\begin{algorithm}[t]
	\caption{Intersecting Blocking Keys}
	\label{alg:intersectKeys}
	\begin{algorithmic}[1]
		\Require{$K$, a dataset of $rid$ to blocks, $b_{0..n}$, where $b_i$ is a 2-tuple of the block key hash, $b.key_i$, and the count of records in this block $b.size_i$}
		\Ensure{$R$, a dataset of $rid$ to intersected blocks, $b_{0..n}$, where $b_i$ is a 2-tuple of the block key hash, $b.key_i$, and the count of records in the \textit{parent} block $b.psize_i$}
		\Function{IntersectKeys}{$K$}
		\State $K \gets \{ (rid, b_{0..n}) \mid (rid, b_{0..n}) \in K \land n \leq \texttt{MAX\_KEYS} \}$ \label{alg:discardMaxKeys}
		\State $R \gets \emptyset$
		\ParFor{$(rid, b_{0..n}) \in K$}
		\LineComment{intersect all block keys, producing $n \choose 2$ new block keys}
		\State $P \gets \{ b_i, b_j \mid b_i \in b, b_j \in b \land b_i < b_j \}$ \label{alg:computeIntersectionsStart}
		\ParFor{$(b_i, b_j) \in P$}
		\State $x.key \gets \Call{Murmur3}{b.key_i, b.key_j}$
		\LineComment{the new block's size is unknown at this point, but we carry the smallest parent's size}
		\State $x.psize \gets min(b.size_i, b.size_j)$ \label{alg:setSmallestParent}
		\State $R[rid] \pluseq x$ \label{alg:computeIntersectionsEnd}
		\EndParFor
		\EndParFor
		
		\State \Return{$R$}
		\EndFunction
	\end{algorithmic}
\end{algorithm}

Algorithm~\ref{alg:intersectKeys} shows how we compute what is semantically a pair-wise intersection of every over-sized block by local operations. The inverted index of blocking keys, $K$, accepted by \method{IntersectKeys}, is logically a map of record ID $rid$ to the set of over-sized blocking keys $b_{0..n}$ for that record. The blocking keys here are represented as a 2-tuple of $(b.key_i, b.size_i)$, where $b.key_i$ is the blocking key hash value and $b.size_i$ is the number of records that share the blocking key $b.key_i$ (as computed by \method{ExactlyCountAndDedupe}). 

We discard from further processing all records which have more than \maxKeysPerRecord blocking keys (line~\algref{alg:intersectKeys}{alg:discardMaxKeys}) as a guard against a quadratic explosion of keys. The governing hypothesis of dynamic blocking is that, since all blocking keys cover a distinct set of record IDs, this quadratic increase in the number of keys will be counterbalanced by the fact that $\left| A \cap B \right| < \left| A \right| + \left| B \right| $ and thus the intersected blocks will tend to become right-sized.  Furthermore, records which have a large number of keys on iteration $i$ are likely to have participated in many right-sized blocks on iterations prior to $i$.

On lines~\algref{alg:intersectKeys}{alg:computeIntersectionsStart} to \algref{alg:intersectKeys}{alg:computeIntersectionsEnd}, we replace the existing, over-sized block keys with $\binom{n}{2}$ new hashes computed by combining every pair of existing over-sized hashes for that record. 
There are some blocking key intersections which we do \textit{not} want to produce. For example, if a dataset had $4$ nearly identical over-sized blocks, then after the first intersection these $4$ columns would intersect with each other to produce $\binom{4}{2} = 6$ columns, but since these were already over-sized and nearly identical, the intersected columns would be over-sized as well. This quadratic growth of over-sized blocking keys per record would not converge. To avoid this hazard, we apply a \textit{progress heuristic} and only keep blocking key intersections that reduce the size of the resulting blocks by some fraction, \maxSimilarity. This heuristic filter is applied in Algorithm~\ref{alg:roughOversizeDetection}, using the minimum \textit{parent} block size which we propagate on line~\algref{alg:intersectKeys}{alg:setSmallestParent}.

\subsubsection{Rough Over-sized Block Detection}

\begin{algorithm}[t]
	\caption{Rough Over-sized Block Detection}
	\label{alg:roughOversizeDetection}
	\begin{algorithmic}[1]
		\Require{$K$, a dataset of $rid$ to over-sized blocks, $b_{0..n}$ where $b_i$ is a 2-tuple of the block key hash, $b.key_i$, and the count of records in the parent block, $b.psize_i$}
		\Ensure{$K_R$, a dataset of record to right-sized blocks}
		\Ensure{$\tilde{K}_O$, a map of record to \textit{possibly} over-sized blocks}
		\Function{RoughOversizeDetection}{$K$}
		\State $cms \gets \Call{ApproxCountBlockingKeys}{K}$ \label{alg:computeCms}
		\State $K_R \gets \emptyset$
		\State $\tilde{K}_O \gets \emptyset$
		\ParFor{$(rid, b_{0..n}) \in K$}
		\ForAll{$b_i \in b$}
		\State $s \gets cms[b.key_i]$
		\State $p \gets b.psize_i$ 
		\If{$s \le \texttt{MAX\_BLOCK\_SIZE}$}
		\State $K_{R}[rid] \pluseq b_i$ \Comment{right-sized}
		\ElsIf{$(s / p) \le \texttt{MAX\_SIMILARITY}$}
		\State $\tilde{K}_{O}[rid] \pluseq b_i$ \Comment{over-sized}
		\EndIf
		\LineComment{We discard over-sized blocks that are too similar in size to parent}
		\EndFor
		\EndParFor
		\State \Return{$K_R, \tilde{K}_O$}
		\EndFunction
	\end{algorithmic}
\end{algorithm}

It is critical that we can accurately count the size of each block. A na\"ive approach would be to pivot from our inverted index to a view of records per blocking key, at which point counting block sizes is trivial. This requires expensive global \textit{shuffling} of all of the data across the cluster in each iteration of the blocking algorithm. Our approach makes novel use of a Count-Min Sketch (CMS) \cite{cormode2005improved} data structure to compute an approximate count of the cardinality of every candidate blocking key (line~\algref{alg:roughOversizeDetection}{alg:computeCms}). We compute one CMS per data partition and then efficiently merge them together.
%
Due to the semantics of a CMS, the approximate count will never be \textit{less} than the true count, and thus no truly over-sized blocks can be erroneously reported as right-sized. In this way, the CMS acts as a filter, dramatically reducing the number of candidate blocks that we need to focus on in each iteration.

\subsubsection{Exactly Count and Deduplicate}
\label{exactlyCountAndDedupe}

\begin{algorithm}[t]
	\caption{Correct over-counting and deduplicate blocks}
	\label{alg:exactCountAndDedupe}
	\begin{algorithmic}[1]
		\Require{$\tilde{K}_{O}$, a dataset of $rid$ to $b_{0..n}$ where $b_i$ are block key hashes for $r$}
		\Ensure{$\hat{K}_R$, a dataset of record to right-sized blocks that were erroneously over-counted by the Count-min Sketch}
		\Ensure{$K_O$, a dataset of truly over-sized blocks, $rid$ to $b_{0..n}$ where $b_i$ is a 2-tuple of the block key hash, $b.key_i$, and the count of records in this block, $b.size_i$}
		\Function{ExactlyCountAndDedupe}{$\tilde{K}_{O}$}
		\State $\hat{K}_R \gets \emptyset$
		\State $K_O \gets \emptyset$
		\State $H \gets \Call{CountKeysAndXORids}{\tilde{K}_{O}}$
		\LineComment{$H$ is dataset of block key hash to tuple of (XOR, size)}
		\State $H_O \gets \{ h \mid h \in H \land h.size > \texttt{MAX\_BLOCK\_SIZE} \}$
		\State $H_U \gets \Call{DropDuplicates}{H_O}$ \Comment{based on XOR} \label{alg:dropDups}
		\LineComment{$counts$ is a broadcasted map of deduped, true over-sized counts}
		\State $counts \gets \Call{BroadcastCounts}{H_U}$ 
		\State $bloom \gets \Call{BuildBloomFilter}{H_O}$ \label{alg:buildBloom}
		
		\ParFor{$(rid, b_{0..n}) \in \tilde{K}_{O}$}
		\ForAll{$b_i \in b$}
		\If{$b.key_i \notin bloom$}
		\State $\hat{K}_{R}[rid] \pluseq b_i$ \Comment{over-counted}
		\ElsIf{$b.key_i \in counts$}
		\State $x.key \gets b.key_i$
		\State $x.size \gets counts[b.key_i]$
		\State $K_{O}[rid] \pluseq x$ \Comment{over-sized}
		\EndIf
		\LineComment{keys in $bloom$ but not in $counts$ were duplicate over-sized blocks, which we discard}
		\EndFor
		\EndParFor
		
		\State \Return{$\hat{K}_R, K_O$}
		\EndFunction
	\end{algorithmic}
\end{algorithm}

Algorithm~\ref{alg:exactCountAndDedupe} focuses on correcting the possibly over-sized blocks. The goal of this method is to partition the $\tilde{K}_O$ blocking keys in the inverted index into three sets, illustrated in Figure~\ref{fig:hdbDiagram}: (1) \textbf{right-sized blocks} that were erroneously over-counted by the Count-Min Sketch, $\hat{K}_R$, which we subsequently union into this iteration's right-sized blocks (line~\algref{alg:hashedDynBlockingTopLevel}{alg:unionCorrectedBlocks}); (2) \textbf{duplicate} over-sized blocks, which we discard; (3) \textbf{surviving, deduplicated} over-sized blocks, $K_O$, with precise counts of how many records are in each, which are then further intersected in the next iteration.

Block $A$ duplicates block $B$ if block $A$'s record IDs are equal to block $B$'s. We arbitrarily discard duplicate blocks, leaving only a single surviving block from the group of duplicates, in order to avoid wasting resources on identical blocks that would only continue to intersect with each other, but produce no new pair-wise comparisons.
%
We do this exact count and dedup in parallel in one map-reduce style operation. To \textit{deduplicate} the blocks we build a \textit{block membership} hash key by hashing each record ID in the candidate block and bit-wise \texttt{XOR}ing them together. Since \texttt{XOR} is commutative, the final block membership hash key is then formed (reduced) by \texttt{XOR}ing the partial membership hash keys.

On line~\algref{alg:exactCountAndDedupe}{alg:dropDups}, we discard duplicate copies of blocking keys that have the same block membership hash key. From these deduplicated blocking keys, $H_U$, we create a string multiset, $counts$, to precisely count the over-sized blocking keys. Even in our largest dataset of over 1 billion records, the largest count of oversized blocks in a particular iteration after deduplication is $\approx$2.6M which easily fits into memory, but if this memory pressure became a scaling concern in the future, we could use another Count-Min Sketch here.

Lastly, we need to distinguish the erroneously over-counted blocks which are actually right-sized, $\hat{K}_R$, from the surviving, deduplicated blocks, $H_U$. On line~\algref{alg:exactCountAndDedupe}{alg:buildBloom} we build a Bloom filter \cite{bloom1970space} over \textit{all} of the over-sized blocking keys, $H_O$, which contains both duplicate and surviving over-sized blocks as determined by precise counting. Therefore, the Bloom filter answers the set membership question: is this blocking key possibly over-sized? In this way, we use this filter as a mechanism to detect right-sized blocks that were erroneously over counted. We build the Bloom filter using a large enough bit array to ensure a low expected false positive rate of $1\mathrm{e}{-8}$. Even in our largest dataset, the biggest Bloom filter that we have needed is less than $\approx$100MB.


%
%

\subsubsection{Pair Deduplication}
\label{pairDedupe}

The final set of right-sized blocks determined after $k$ iterations of Hashed Dynamic Blocking will likely contain blocks that overlap or are entirely subsumed by other blocks. 
We use a map-reduce sequence to compute all distinct pairs similar to the pair deduplication algorithm presented in \cite{McNeill2012}, retaining only the pair from the \textit{largest} block in the case of duplicates. 
This results in tuples $(rid_1, rid_2, b.key_i)$ where $b.key_i$ is the identifier for the \textit{largest} block that produced the pair $(rid_1, rid_2)$. We then group the tuples by $b.key_i$ to reconstruct the blocks. For each block, we now have an edgelist of $[1, \binom{n}{2}]$ pairs and have the complete set of $n$ resulting record IDs. We build a bitmap of $\binom{n}{2}$ bits with each representing one pair for pairwise matching. The bit index $b_{i,j}$ for a pair of record IDs $\text{rec}_a$, $\text{rec}_b$ is computed by: $b_{i,j} = i * (n - 1) - (i - 1) * i / 2 + j - i - 1$, where $i, j$ are the zero-based \textit{indexes} of $\text{rec}_a$, $\text{rec}_b$ in the block of $n$ records, ordered by the record IDs natural order and $i < j$. This is simply a sequential encoding of the strictly upper triangular matrix describing all $\binom{n}{2}$ pairs in the block. 
In the common case where none of the $\binom{n}{2}$ pairs are filtered out, we omit the bitmap and just score all pairs from the block during pairwise matching.

\section{Prior Work}

\subsection{Prior work on Dynamic Blocking}
The need for Hashed Dynamic Blocking may be unclear since its semantics (the pairs produced after pair deduplication) are essentially the same as that of \cite{McNeill2012} for scalar-valued attributes.  Relative to \cite{McNeill2012}, this work offers the following advantages:
(1) \cite{McNeill2012} had a substantial memory and I/O footprint since the content of the records being blocked had to be carried through each iteration of the algorithm.
%
(2) LSH would have been challenging to implement in the Dynamic Blocking algorithm of \cite{McNeill2012} as it did not contemplate blocking on array-valued columns. 

\subsection{Meta-Blocking based approaches}
\label{metablocking}

%
Meta-Blocking \cite{Papadakis2014}, like dynamic blocking, starts from an input collection of blocks and is independent of the scheme for generating these blocks.  It encompasses a broad variety of techniques, but at its most basic, it builds a graph in which a node corresponds to a record and an edge $(e_1,e_2)$ indicates that at least one block contains both $e_1$ and $e_2$.  

A shortcoming of the Meta-Blocking family of algorithms is that it is linear in the number of comparisons in the input block collection \cite{Efthymiou2017}, which would be equivalent to the total number of comparisons implied by all blocks (both \emph{oversized} and \emph{right-sized}) in the input block collection.  Meta-Blocking approaches generally mitigate this linearity by purging the very largest blocks \cite{Efthymiou2017} and by Block Filtering \cite{Papadakis2016a} which, for each entity, trims the entity from the largest blocks in which it participates.  However, \cite{Efthymiou2017} reports only an ``at least 50\%'' reduction in the number of pairwise comparisons using Block Filtering, leaving the algorithm still linear in the comparisons of the input block collection. 

Our approach, by contrast, aims to leverage rather than trim large blocks by intersecting them with other large blocks.  This is better than either discarding or trimming the large blocks, which may sacrifice recall, or attempting to do even a minimal amount of pairwise processing on the large blocks, which would impact performance.  

BLAST \cite{simonini2019scaling} is a schema-aware meta-blocking approach.  One innovation of BLAST is making records that share high entropy attributes like \emph{name} more likely to be pairwise-compared than low entropy attributes like \emph{year of birth}.  HDB, as noted above, takes this approach one step further by making rare values (\eg surname \emph{Fishburne}) more likely to create pairs than common values (\eg surname \emph{Jones}). 

\section{Experimental Results}
\label{experiments}

\begin{table*}[t]
    \centering
    \caption{Datasets used for experiments where $BB$ indicates (L)SH or (T)oken block building strategy and positive labels marked with $\dagger$ are complete ground truth. Datasets marked $C$ are Commercial datasets.}
    \begin{subfigure}[t]{0.45\textwidth}
		\resizebox{\linewidth}{!}{
		\begin{tabular}{|Hl|r|c|c|c|c|}
			\hline
			Dataset & Moniker & Records & +Labels & Cols & $BB$ & Src \\
			\hline
			Variation 1M & VAR1M & 1.03M & 818 & 60 & L & C \\
			Variation 10M & VAR10M & 10.36M & 8,890 & 60 & L & C \\
			Variation 25M & VAR25M & 25.09M & 20,797 & 60 & L & C \\
			Variation 50M & VAR50M & 50.02M & 40,448 & 60 & L & C \\
			Variation 107M & VAR107M & 107.58M & 80,068 & 60 & L & C \\
			Variation 530M & VAR530M & 530.73M & 76,316 & 60 & L & C \\
			\hline 
		\end{tabular} 	
		}
    \end{subfigure}
    \quad
    \begin{subfigure}[t]{0.45\textwidth}
		\resizebox{\linewidth}{!}{
		\begin{tabular}{|Hl|r|c|c|c|c|}
			\hline
			Dataset & Moniker & Records & +Labels & Cols & $BB$ & Src \\
			\hline
			Ohio Voter & VOTER & 4.50M & 53,653$^\dagger$ & 108 & L & \cite{ohioVoterWebsite} \\
			DBLP-Scholar & SCHOLAR & 64,263 & 7,852$^\dagger$ & 5 & L & \cite{kopcke} \\
			DBLP-Citeseer & CITESR & 4.33M & 558k$^\dagger$ & 7 & L & \cite{simonini2019scaling} \\
			DBpedia & DBPEDIA & 3.33M & 891k$^\dagger$ & --- & T & \cite{Efthymiou2017} \\
			Freebase & FREEB & 7.11M & 1.31M$^\dagger$ & --- & T & \cite{Efthymiou2017} \\
			\hline 
		\end{tabular} 	
		}
	\end{subfigure}
    \label{tbl:datasets}
\end{table*}


We present experimental results to explore a few different aspects of Hashed Dynamic Blocking: (1) we present metrics illustrating the overall performance of HDB on a diverse collection of datasets compared to two different baselines: (a) Threshold Blocking (THR) and (b) Parallel Meta-blocking\footnote{https://github.com/vefthym/ParallelMetablocking} (PMB) \cite{Efthymiou2017}. Threshold Blocking refers to blocking based on field values (as in HDB and PMB), but if a block is too large (records $> 500$) then the block is discarded entirely. This simple baseline is useful in illustrating the value of \textit{dynamic} blocking as a means to discover co-occurring values that are discriminating enough to warrant all pairs comparison. (2) we demonstrate the impact of using LSH-based block building with varying parameter values of $b$ bands and $w$ minhashes per band. Unless mentioned otherwise, we use hyper-parameters $\maxBlockSize = 500$, $\maxKeysPerRecord = 80$, and $\maxSimilarity = 0.9$.

We run all of our experiments using AWS ElasticMapReduce (EMR) using Spark 2.4.3 and 100 m4.4xlarge core nodes with 20GB of executor memory and 16 cores per executor. At the time of writing m4.4xlarge instances on AWS cost \$1.04/hour (on EMR). Our largest dataset of 530M records takes 169 minutes to complete blocking costing $\approx$\$307. 

\subsection{Datasets}
\label{datasets}

In order to evaluate the effectiveness of Hashed Dynamic Blocking, we evaluate against a diverse collection of datasets. 
Table~\ref{tbl:datasets} lists summary information for each dataset used for evaluation. The \textbf{VARxx} datasets are product variations datasets that come from a subset of a large product catalog from a large E-commerce retailer. Each \textbf{VARxx} record contains sparsely populated fields such as product name, description, manufacturer, and keywords. 
%
We include the bibliographic citation dataset \textit{DBLP-Scholar}, called \textbf{SCHOLAR}, from \cite{kopcke} as it is small enough to practically illustrate the differences between LSH-based block building with many different configurations. We include the DBLP-Citeseer bibliographic citation dataset, \textbf{CITESR}, from \cite{simonini2019scaling}. 
We also include two token blocking-based datasets, \textbf{DBPEDIA} and \textbf{FREEB} (Freebase), published in \cite{Efthymiou2017}. For these two datasets, instead of using our LSH-based block building method, we use the exact token-blocking input published by the authors in \cite{Efthymiou2017}.

Finally, we introduce a large, labeled, public-domain dataset: the Ohio Voter dataset, called \textbf{VOTER}. We built this by downloading two snapshots of Ohio registrations \cite{ohioVoterWebsite} from different points in time and treating as duplicates records with the same voter ID but different demographic information.
Similar work was done previously on North Carolina voter registration data  \cite{christen2013preparation}, but the Ohio dataset is richer in that it contains 108 columns, including birthday and voter registration.

\subsection{Metrics}
We present a few different metrics in order to evaluate performance. We use the established $PQ$, Pair Quality, (analogous to precision) and $PC$, Pair Completeness, (analogous to recall) metrics \cite{Christen2011}. Every dataset described in Table~\ref{tbl:datasets} has labeled pair-wise training data. We report the number of positively labeled pairs as \textit{+Labels}. However, as is a common problem in record linkage evaluation, the number of labeled training pairs is usually incomplete and significantly smaller than the possible pairs in the input record set. 
For these incompletely labeled datasets, we present $PC$ with respect to the labeled pairs, defined as: $ |P \cap L^{+}| / |L^{+}| $, where $P$ is the set of pairs produced by the blocker and $L^{+}$ is the set of positively labeled pairs in the training data.
 
To measure $PQ$ for datasets without complete ground truth, we follow a similar practice as described in \cite{McNeill2012} where we employ an \textit{Oracle} pair-wise model previously trained on the complete labeled dataset. We use the same oracle per dataset for all experiments and thus the numbers are relatively comparable, despite containing some error introduced by the imperfect Oracle. 


\subsection{Comparing Hashed Dynamic Blocking to other methods}

\newcommand{\resultscolwidth}{1.3cm} 
\newcommand{\timecolwidth}{0.7cm} 
\begin{table*}[t]
	\centering
	\caption{Comparing Hash Dynamic Blocking (HDB) to other methods based on Pair Quality (PQ), Pair Completeness (PC), and Elapsed Time (T) in minutes}
	\begin{tabular}{%
	|p{1.5cm}|>{\centering\arraybackslash}p{\resultscolwidth}|>{\centering\arraybackslash}p{\resultscolwidth}|>{\centering\arraybackslash}p{\timecolwidth}|>{\centering\arraybackslash}p{\resultscolwidth}|>{\centering\arraybackslash}p{\resultscolwidth}|>{\centering\arraybackslash}p{\timecolwidth}|>{\centering\arraybackslash}p{\resultscolwidth}|>{\centering\arraybackslash}p{\resultscolwidth}|>{\centering\arraybackslash}p{\timecolwidth}|}
	\hline
	& \multicolumn{3}{c|}{Threshold (THR)} & \multicolumn{3}{c|}{Parallel Meta (PMB)} & \multicolumn{3}{c|}{Hash Dynamic (HDB)} \\ 
	\hline
	Dataset & $PQ$ & $PC$ & $T$ & $PQ$ & $PC$ & $T$ & $PQ$ & $PC$ & $T$ \\ 
	\hline

VAR1 & \textbf{0.3003} & \textbf{0.9450} & 2.6 & 0.1825 & 0.5428 & 13.9 & 0.2612 & \textbf{0.9450} & 4.9 \\
VAR10 & \textbf{0.3336} & 0.9444 & 4.8 & 0.1174 & 0.7773 & 16.3 & 0.3083 & \textbf{0.9445} & 10.6 \\
VAR25 & \textbf{0.3445} & 0.9251& 7.8 & 0.1213 & 0.7620 & 23.8 & 0.2739 & \textbf{0.9315} & 20.7 \\
VAR50 & \textbf{0.3355} & 0.9240 & 13.9 &  &  &   & 0.2394 & \textbf{0.9343} & 30.5 \\
VAR107 & \textbf{0.3227} & 0.9102 & 23.8 &  &  &   & 0.2168 & \textbf{0.9277} & 54.6 \\
VAR530 & 0.4787 & 0.8341 & 110.3 &  &  &   & \textbf{0.4834} & \textbf{0.8588} & 169.2 \\
VOTER & \boldmath{$\nexpnum{6.96}{4}$} & \textbf{1.0000} & 2.6 & $\nexpnum{2.74}{4}$ & 0.9986 & 24.2 & $\nexpnum{5.19}{4}$ & \textbf{1.0000} & 5.4 \\
SCHOLAR & \boldmath{$\nexpnum{5.52}{3}$} & \textbf{0.4749} & 0.8 & $\nexpnum{1.71}{3}$ & 0.3583 & 10.5 & $\nexpnum{5.52}{3}$ & \textbf{0.4749} & 1.2 \\
CITESR & \boldmath{$\nexpnum{1.32}{2}$} & 0.9544 & 2.2 & $\nexpnum{5.02}{4}$ & 0.4808 & 15.5 & $\nexpnum{5.58}{3}$ & \textbf{0.9545} & 3.6 \\
DBPEDIA & \boldmath{$\nexpnum{7.99}{4}$} & 0.9376 & 3.6 & $\nexpnum{2.60}{4}$ & 0.9742 & 14.5 & $\nexpnum{2.38}{4}$ & \textbf{0.9921} & 22.1 \\
FREEB & \boldmath{$\nexpnum{2.37}{4}$} & 0.7340 & 6.8 & $\nexpnum{1.50}{4}$ & 0.8303 & 23.8 & $\nexpnum{1.47}{4}$ & \textbf{0.8497} & 25.9 \\

		\hline
	\end{tabular}
	\label{idbRuns}
\end{table*}



Table~\ref{idbRuns} shows the performance of Hashed Dynamic Blocking (HDB). Threshold Blocking is a simple strategy, but comparing the results to HDB shows that in all large datasets, recall is hurt when we simply discard over-sized blocks. 
Table~\ref{pairCountsPerAlgo} shows the number of pairs produced in each of our experiment setups. Since PMB and HDB have different hyper-parameters that affect the operating point, we configured both through trial and error to produce a similar number of pairs to relatively evaluate $PC$ and $PQ$ at the same operating point. \textit{Naive} here represents the number of pairs produced by simple blocking on the blocking key values; that is comparing all pairs of records that share any blocking key value and not discarding any blocks due to size (as we do in THR). The VAR530 dataset with Naive blocking produces 120 quadrillion pairs, which on our cluster would take over $7,200$ years to score, highlighting the need for more sophisticated blocking approaches at massive scale.

We have not been able to successfully execute PMB on some of the larger sets (VAR50, VAR107, VAR530); it fails with out-of-memory errors and we have been unable to get it to complete. We ran into similar issues when running BLAST \cite{simonini2019scaling} on our huge datasets, which we expected given that they broadcast hash maps of record ID $\to$ blocking keys to every node. For our large datasets, this single broadcast map would be multiple TBs of memory.

We note that HDB demonstrates improved recall over PMB despite PMB producing more pairs to evaluate. We believe this may be a consequence of the heuristic of meta-blocking weighting pairs that occur in multiple blocks. In the case of LSH-based blocking keys where there are many highly overlapping blocks, this may result in PMB picking many redundant pairs that don't improve compression. HDB by contrast, prefers to focus on the blocks that are small enough to thoroughly evaluate and find intersections of over-sized blocks. This may produce more diversity in the pairs emitted by HDB compared to PMB.

\newcommand{\cmpColWid}{0.93cm}
\begin{table*}[t]
    \centering
    \caption{Comparing the number of pairs produced by different blocking algorithms}
    \begin{subfigure}[t]{0.45\textwidth}
		\resizebox{\linewidth}{!}{
		\begin{tabular}{%
		|p{1.7cm}|>{\centering\arraybackslash}p{\cmpColWid}|>{\centering\arraybackslash}p{\cmpColWid}|
		>{\centering\arraybackslash}p{\cmpColWid}|>{\centering\arraybackslash}p{\cmpColWid}|%
		}
			\hline
			& Naive & THR & PMB & HDB  \\ 
			\hline
			Dataset & $||B||$ & $||B||$ & $||B||$ & $||B||$ \\ 
			\hline
			VAR1 & $\expnum{4.5}{11}$ & $\expnum{1.1}{8}$ & $\expnum{1.8}{8}$ & $\expnum{1.6}{8}$ \\
			VAR10 & $\expnum{4.4}{13}$ & $\expnum{1.1}{9}$ & $\expnum{1.7}{9}$ & $\expnum{1.4}{9}$ \\
			VAR25 & $\expnum{2.6}{14}$ & $\expnum{3.2}{9}$ & $\expnum{5.4}{9}$ & $\expnum{5.3}{9}$ \\
			VAR50 & $\expnum{1.1}{15}$ & $\expnum{6.4}{9}$ &  & $\expnum{1.3}{10}$ \\
			VAR107 & $\expnum{5.2}{15}$ & $\expnum{1.4}{10}$ &  & $\expnum{3.0}{10}$ \\
			VAR530 & $\expnum{1.2}{17}$ & $\expnum{4.5}{10}$ &  & $\expnum{6.8}{10}$ \\
			\hline
		\end{tabular}
		}
	\end{subfigure}
    \quad
    \begin{subfigure}[t]{0.45\textwidth}
		\resizebox{\linewidth}{!}{
		\begin{tabular}{%
		|p{1.7cm}|>{\centering\arraybackslash}p{\cmpColWid}|>{\centering\arraybackslash}p{\cmpColWid}|
		>{\centering\arraybackslash}p{\cmpColWid}|>{\centering\arraybackslash}p{\cmpColWid}|%
		}
			\hline
			& Naive & THR & PMB & HDB  \\ 
			\hline
			Dataset & $||B||$ & $||B||$ & $||B||$ & $||B||$ \\ 
			\hline
			VOTER & $\expnum{3.5}{11}$ & $\expnum{9.3}{8}$ & $\expnum{2.3}{9}$ & $\expnum{1.3}{9}$ \\
			SCHOLAR & $\expnum{2.4}{7}$ & $\expnum{2.0}{6}$ & $\expnum{4.7}{6}$ & $\expnum{2.0}{6}$ \\
			CITESR & $\expnum{2.3}{11}$ & $\expnum{4.5}{7}$ & $\expnum{6.2}{8}$ & $\expnum{1.1}{8}$ \\
			DBPEDIA & $\expnum{8.0}{10}$ & $\expnum{1.0}{9}$ & $\expnum{3.2}{9}$ & $\expnum{3.7}{9}$ \\
			FREEB & $\expnum{2.2}{11}$ & $\expnum{4.1}{9}$ & $\expnum{7.0}{9}$ & $\expnum{7.6}{9}$ \\
			\hline
		\end{tabular}
		}
	\end{subfigure}
	\label{pairCountsPerAlgo}
\end{table*}

%
%

\subsection{Comparing LSH Configurations}
\label{lshExperiment}

To illustrate the impact of including Locality Sensitive Hashing (LSH) with HDB we present numbers showing how $PQ$ and $PC$ are affected by various LSH configurations. Figure~\ref{lshResultsCharts} shows the results with varying the number of \textit{bands} $b$ between 3 and 16 and varying the number of minhashes per band $w$ from 8 to 3. As expected, LSH improves recall for datasets that have multi-token text fields, which is most of the datasets evaluated. In some instances, adding LSH dramatically improves recall. As expected, for most datasets the precision decreases as LSH becomes more liberal. Figure~\ref{lshScholarScatter} shows a scatter plot of many different LSH configurations on the SCHOLAR dataset and includes the Token Blocking ($HDB  TB=0.5$) result for comparison to highlight the differences in $PQ$ and $PC$. The diameter of each data point in the scatter plot is a linear scaling of the number of pairs produced.

\newcommand{\lshmarksize}{3 pt}
\newcommand{\lshmarksizepq}{4 pt}
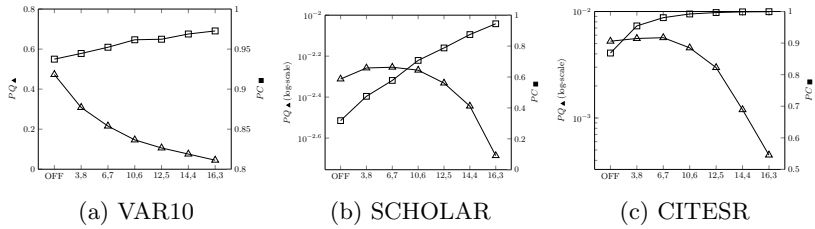
\begin{figure*}[t]
    \centering
    \begin{subfigure}[b]{0.29\textwidth}

		\resizebox{\linewidth}{!}{
		\begin{tikzpicture}
			
		  \begin{axis}[ymin=0.0
		  	,ymax=0.8
		  	,xtick=data
		  	,xticklabels={OFF,{3,8},{6,7},{10,6},{12,5},{14,4},{16,3}}
		  	,ylabel={$PQ$ \myplotmark{triangle*}}
		  	,axis y line*=left
		  	,ylabel near ticks
		  	]
		    \addplot[mark=triangle
		    	,discard if not={dataset}{VAR10}
		    	,mark size=\lshmarksizepq
		    	] table[x=lshxaxis, y=pq, col sep=comma] {data/lsh-results.csv};
		  \end{axis}
		  \begin{axis}[ymin=0.8
		  		,ymax=1.0
      			,ylabel={$PC$ \myplotmark{square*}}
      			,axis y line*=right
      			,axis x line=none
      			,ylabel near ticks
      		]
		    \addplot[mark=square,mark size=\lshmarksize,discard if not={dataset}{VAR10}] table[x=lshxaxis, y=pc, col sep=comma] {data/lsh-results.csv};
		  \end{axis}
		\end{tikzpicture}
		}
		
		\label{lshResultsChartVar25}
        \caption{VAR10}
    \end{subfigure}
    \begin{subfigure}[b]{0.29\textwidth}
		
		\resizebox{\linewidth}{!}{
		\begin{tikzpicture}
		  \begin{axis}[ymin=0.0
		  	,ymax=0.01
		  	,xtick=data
		  	,xticklabels={OFF,{3,8},{6,7},{10,6},{12,5},{14,4},{16,3}}
		  	,ylabel={$PQ$ \myplotmark{triangle*} (log-scale)}
		  	,axis y line*=left
		  	,ylabel near ticks
		  	,yminorticks=true
		  	,ymode=log
       		,log basis y={10}
		  	]
		    \addplot[mark=triangle,mark size=\lshmarksizepq,discard if not={dataset}{SCHOLAR}] table[x=lshxaxis, y=pq, col sep=comma] {data/lsh-results.csv};
		   \end{axis}
		   \begin{axis}[ymin=0.0
		  		,ymax=1.0
      			,ylabel={$PC$ \myplotmark{square*}}
      			,axis y line*=right
      			,axis x line=none
      			,ylabel near ticks
      		]
		    \addplot[mark=square,mark size=\lshmarksize,discard if not={dataset}{SCHOLAR}] table[x=lshxaxis, y=pc, col sep=comma] {data/lsh-results.csv};
		  \end{axis}
		\end{tikzpicture}
		}

		\label{lshResultsScholar}
        \caption{SCHOLAR}
    \end{subfigure}
    \begin{subfigure}[b]{0.29\textwidth}
		
		\resizebox{\linewidth}{!}{
		\begin{tikzpicture}
		  \begin{axis}[ymin=0.0
		  	,ymax=0.01
		  	,xtick=data
		  	,xticklabels={OFF,{3,8},{6,7},{10,6},{12,5},{14,4},{16,3}}
		  	,ylabel={$PQ$ \myplotmark{triangle*} (log-scale)}
		  	,axis y line*=left
		  	,ylabel near ticks
		  	,yminorticks=true
		  	,ymode=log
       		,log basis y={10}
		  	]
		    \addplot[mark=triangle,mark size=\lshmarksizepq,discard if not={dataset}{CITESR}] table[x=lshxaxis, y=pq, col sep=comma] {data/lsh-results.csv};
		  \end{axis}
		  \begin{axis}[ymin=0.5
		  		,ymax=1.0
      			,ylabel={$PC$ \myplotmark{square*}}
      			,axis y line*=right
      			,axis x line=none
      			,ylabel near ticks
      		]
		    \addplot[mark=square,mark size=\lshmarksize,discard if not={dataset}{CITESR}] table[x=lshxaxis, y=pc, col sep=comma] {data/lsh-results.csv};
		  \end{axis}
		\end{tikzpicture}
		}

		\label{lshResultsCitesr}
        \caption{CITESR}
    \end{subfigure}
    \caption{$PQ$ and $PC$ of various $\text{LSH}(b,w)$ settings on three datasets with text fields}
    \label{lshResultsCharts}
\end{figure*}

\section{Conclusions}


We have shown Hashed Dynamic Blocking being applied to different large datasets up to 530M records. We also introduced the LSH-based block building technique, and illustrated its usefulness in blocking huge datasets. The Hashed Dynamic Blocking algorithm leverages a fortunate convergence in the requirements for efficiency and accuracy. 
%
%
HDB accomplishes this through a new algorithm which iteratively intersects and counts sets of record IDs using an inverted index and approximate counting and membership data structures. This efficient implementation is fast, robust, cross-domain, and schema-independent, thus making it an attractive option for blocking large complex databases.



\bibliographystyle{splncs04}
\bibliography{FastDynamicBlocking}  

\end{document}